\documentclass[twocolumn,showpacs,preprintnumbers,amsmath,amssymb,prl,floatfix]{revtex4-1}

\usepackage{graphicx}
\usepackage{dcolumn}
\usepackage{bm}
\usepackage[tight]{subfigure}
\usepackage{amsmath}
\usepackage{verbatim}
\usepackage{color}
\newcommand{\fracs}[2]{{\textstyle \frac{#1}{#2}}}  
\newcommand{\ket}[1]{|{#1}\rangle}
\renewcommand{\vec}[1]{\mathbf{#1}}

\newcommand{\Fig}[1]{Fig.~\ref{#1}}
\newcommand{\Eq}[1]{Eq.~(\ref{#1})}

\newcommand{\ham}[1]{H_\text{#1}}
\newcommand{\aks}{a_{{k}\sigma}^{\phantom{\dagger}}}
\newcommand{\aksd}{a_{{k}\sigma}^{\dagger}}
\newcommand{\bd}{b_{\phantom \sigma}^{\dagger}}
\newcommand{\akpsp}{a_{{k'}\sigma'}^{\phantom{\dagger}}}
\newcommand{\epsi}{\epsilon_{{k}\sigma}^{\phantom{\dagger}}}
\newcommand{\HK}{high-spin Kondo~}
\newcommand{\QK}{QST Kondo~}
\newcommand{\VQK}{vibration-induced QST Kondo~}

\newcommand{\cut}[1]{}

\begin{document}
\title{
  Interaction of spin and vibrations in transport through single-molecule magnets
}
\author{F. May$^{(1,2)}$}
\author{M. R. Wegewijs$^{(3,4)}$}
\author{W. Hofstetter$^{(1)}$}
\affiliation{
(1) Institut f\"ur Theoretische Physik, Johann Wolfgang Goethe-Universit\"at, 60438 Frankfurt/Main, Germany}
\affiliation{
(2) Max Planck Institute for Polymer Research, 55128 Mainz, Germany}
\affiliation{
(3) Institut f\"ur Theorie der Statistischen Physik, RWTH Aachen, 52056 Aachen,  Germany}
\affiliation{
(4) Peter Gr{\"u}nberg Institut and JARA - Fundamentals of Information Technology,
 Forschungszentrum J{\"u}lich, 52425 J{\"u}lich,  Germany }
\begin{abstract}
We study linear electron transport through a single-molecule magnet (SMM)
and the interplay of its anisotropic spin with quantized vibrational distortions of the molecule.
We show that, despite the longitudinal anisotropy barrier and small transverse anisotropy,
vibrational fluctuations can induce quantum spin-tunneling (QST) and a QST-Kondo effect.
The interplay of spin scattering, QST and  molecular vibrations can strongly enhance the Kondo effect
and induce an anomalous magnetic field dependence of vibrational Kondo side-bands.
\end{abstract}

\maketitle
Transport measurements on nanometer-sized magnetic systems address the fundamental problem
of how a few magnetic atoms in an anisotropic environment respond to an electron current~\cite{Bogani08}.
Such an environment is provided, for instance,
 by ligand groups holding such atoms together in a single magnetic molecule contacted in a break junction~\cite{Zyazin10,Parks10}.
A very similar situation arises for transport through magnetic atoms embedded in a molecular network on an insulating surface in an STM setup~\cite{Otte08,*Otte09}.
Such systems, for simplicity all referred to as single-molecule magnets (SMM), constitute a single, large spin-moment with spin-anisotropy.
The interplay with quantum transport provides new possibilities to study and control their  molecular magnetism.
For instance, the tunneling allows access to several charge states of the SMM which can exhibit enhanced magnetic properties~\cite{Zyazin10}.
When such charge states are only virtually accessible, effective spin-spin exchange interaction arises~\cite{Otte08,*Otte09} and inelastic excitation of the spin moment is possible~\cite{Zyazin10}, allowing for time-dependent control~\cite{Loth10}.
A key result is that in either regime the transport depends sensitively on the magnetic anisotropy of the SMM, which is characterized by spin-quadrupole terms in the Hamiltonian.
A further new aspect is the recently demonstrated mechanical tuning of these anisotropy terms in a transport setup~\cite{Parks10}.
Although the effect of such molecular distortions on magnetization measurements of SMM crystal samples has been addressed~\cite{Leuenberger00,*Pohjola00,Kortus02},
so far their dynamic effect on transport through an SMM have not been studied.
One candidate for sensitively probing such a coupling of the SMM \emph{spin} to vibrations is a specific type of Kondo effect induced by quantum spin-tunneling (QST).
This QST through the energy barrier arising from a dominant uni-axial magnetic anisotropy term
relies on  the presence of weak transverse anisotropy.
Combined with the exchange tunneling to attached electrodes
a QST-Kondo resonance specific to SMMs can arise~\cite{Romeike06a}.
One  might expect such QST-assisted Kondo transport to be simply suppressed
by coupling of the spin to molecular vibrations as this tends to increase the anisotropy barrier~\cite{Kortus02}.
However, the dynamical effect of vibrational fluctuations and the possible competition between longitudinal and transverse spin-vibration coupling have not been studied so far,
even though coupling to vibrations in the Kondo regime has been considered for spin-isotropic molecules~\cite{Paaske05,Kikoin06,Balseiro06,*Cornaglia07}.

In this paper we consider the modulation of the magnetic anisotropy of an SMM by a
quantized vibrational mode distorting an SMM with half-integer spin.
Strikingly, even \emph{without static transverse anisotropy},
a QST-induced Kondo peak can arise in the linear conductance.
This Kondo  effect is dynamically generated by vibrational fluctuations which distort the SMM,
and thereby allow the spin to fluctuate.
More generally, a higher QST-Kondo temperature may result from spin-vibration coupling
which is relevant for experimental investigation of low temperature transport through SMMs.
This enhancement of the interplay of Kondo spin scattering and QST
by discrete vibrations
indicates a possible avenue along which transport and quantum  magnetism may be combined with nano-mechanical effects.

\begin{figure}[t!]
  \includegraphics[height=0.5\linewidth]{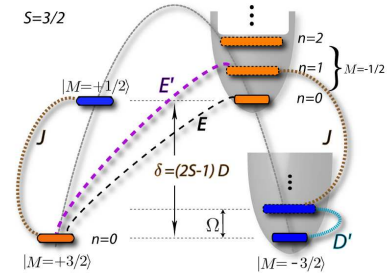}
  \caption{
    \label{fig:1}
    (Color online).
    Magneto-mechanical excitation spectrum of the SMM with $S=3/2$.
    Magnetic states, lying on an inverted parabola  due to the static longitudinal anisotropy ($D$),
    each have vibrational excitations ($n$) on the upright parabolas.
    Magnetic transitions induced by  Kondo spin-scattering ($J$), static anisotropy ($E$)
    and longitudinal ($D'$) and transverse ($E'$) spin-vibration coupling are indicated.
    Magnetic states in the two different Kramer's subspaces are marked blue and orange.
  }
\end{figure}

{\em Model and method.}
We consider a SMM strongly coupled to electronic 
leads at low temperature in the Coulomb blockade regime, where the charge on the SMM only changes virtually.
We assume that the spin couples to a local vibrational mode with frequency $\Omega$.
The total Hamiltonian reads $H=\ham{SMM}+\ham{K}$ with
\begin{align}
\ham{SMM} & = -D S_z^2 +\fracs{1}{2}E \left( S_{+}^2+S_{-}^2\right) +\Omega \bd b \label{eq1}\\
 	  & + \big[ -D' S_z^2 + \fracs{1}{2}E' \left(S_+^2+S_{-}^2 \right) \big] \left(\bd+b\right) \nonumber \\
\ham{K}&=\sum_{k\sigma}\epsi \aksd \aks + J\boldsymbol{S} \cdot \boldsymbol{s} \,. \label{eq2}
\end{align}
Here $S_z$ is the projection of the molecule's spin on its easy axis, which we choose to be the $z$-axis and $S_{\pm}=S_x\pm i S_y$.
We consider here only half-integer values of the spin magnitude $S$, for which there is a Kondo effect at zero magnetic field~\cite{Romeike06a},
and later comment on the integer spin case.
Starting from an isolated molecule,
the longitudinal anisotropy $D$ splits the eigenstates of $S_z$ into the inverted parabolic magnetic spectrum,
which is sketched in \Fig{fig:1} for the representative case of $S=3/2$ used throughout this work.
The zero-field splitting (ZFS) corresponds to the energy difference between the ground-state and the first magnetic excitation, and equals $\delta=(2S-1)D$ for $E=0$ (and $E'=D'=0$).
The transverse anisotropy $E$ breaks the continuous rotational symmetry about the easy-axis of the SMM,
thereby causing spin-tunneling through the barrier.
As shown in~\cite{Kortus02}, vibrational modes modulate the magnetic anisotropy and can significantly contribute to the observed magnetic splittings.
Here we additionally consider the dynamical effects of such coupling by allowing the dominant anisotropy parameters to depend linearly on the vibrational mode coordinate $Q=(b+\bd)/\sqrt{2}$ through coupling coefficients $D'$ and $E'$.
Here the operator $b$ ($b^\dagger$) relaxes (excites) the vibration by one quantum.
Thus, when the SMM vibrates it lowers its symmetry and QST is enhanced.
Importantly, this also holds for virtual quantum vibrations.
We note that recently such a linear dependence of the $D$ parameter on the pitch angle coordinate~\cite{Gregoli09} in the tetra-iron (III) ``propeller''-SMM used in~\cite{Zyazin10} has been measured.
The conduction electron states, represented by the operators $\aks,\aksd$  in~\Eq{eq2}, correspond to even combinations of left and right physical electronic states and their bandwidth is given by $2W$.
Finally, deep in the Coulomb blockade regime, the interaction of the SMM with the electrodes is given by an isotropic Heisenberg spin-exchange with the conduction band electron spin
$\vec{s}=\fracs{1}{2} \sum_{kk'}\sum_{\sigma\sigma'}\aksd \boldsymbol \tau_{\sigma\sigma'}\akpsp$
where $\boldsymbol \tau$ is the vector of Pauli matrices.
The coupling $J$ is assumed anti-ferromagnetic, which, as pointed out in~\cite{Gonzalez08}, depends on the spins of the virtual charge states of the SMM~\cite{Aligia86,*Lustfeld81}.
One might expect that coupling to a molecular vibration suppresses the interaction of the SMM with the electrodes due to Franck-Condon overlap, effectively reducing $J$.
However, for an isotropic spin 1/2 it was shown that the opposite may happen and that spin-exchange processes which change the vibrational quantum number are suppressed deep in the Coulomb blockade regime~\cite{Paaske05}.
Therefore
we assume $J$ to be independent of the vibrational coordinate in~\Eq{eq2}. 

The numerical results for the zero-temperature linear conductance
$G(V) = (e^2/\hbar) \left( A(eV/2) + A(-eV/2) \right)$
 as a function of the bias voltage $V$ presented below were calculated using the Numerical Renormalization Group (NRG).
Here $A(\omega)=\sum_{\sigma} A_\sigma(\omega)$ is the equilibrium SMM spectral function obtained from the T-matrix~\cite{Costi00}.
This method is exact in the limit of linear response in $V$ or of strong asymmetric coupling of the SMM to the electrodes.
For all NRG calculations we used the parameters $\Lambda=2$, $N_s=4000$ states~\cite{Bulla08rev} and 11 vibrational states were sufficient to obtain results independent 
of the vibration number cutoff.

{\em Static anisotropy.}
Even without the vibration ($E'=D'=0$) or transport ($J=0$)
 the SMM eigenstates are not spin-eigenstates when $E \ne 0$.
However, 
for the typical case of moderate transverse anisotropy $E<D$
it still is convenient to label these mixed states by the dominant spin-eigenstate ($M$) in the superposition.
Due to the two-fold spin-rotational symmetry of \Eq{eq1} the mixing caused by $E$ is only possible within the two subspaces spanned by $\ket{\pm \fracs{1}{2}},\ket{\mp \fracs{3}{2}}$ (marked blue and orange in~\Fig{fig:1}).
It was shown~\cite{Romeike06a} that upon including exchange spin scattering with conduction band electrons ($J$) the interplay with the QST (generated by $E$) gives rise to a Kondo peak in the linear conductance.
Due to the presence of the electrodes spin-fluctuations thus become significant at low temperature despite the presence of the anisotropy barrier of size $DS^2$ opposing SMM spin-reversal.
A hallmark of this QST-Kondo effect is that it is suppressed with decreasing ratio of $E/D$ or increasing $S$ (because the barrier grows).
This QST-Kondo effect is clearly distinct from the under-screened high-spin Kondo effect which arises for $S\geq 1$ in the limit $D=E=D'=E'=\Omega=0$ where magnetic anisotropy is not important~\cite{Koller05}.
Starting from the latter limit, introducing the anisotropy barrier, $D>0$, both splits and suppresses the \HK peak.
The remnants of the \HK peak are located close to the ZFS scale $\delta =(2S-1)D$, with possible renormalization to smaller values for large exchange interaction $J$~\cite{Zyazin10}.

These ZFS \HK side-peaks have recently been studied in detail in several experiments~\cite{Zyazin10,Parks10,Otte08,*Otte09}.
The \QK peak, on the other hand, located at zero-bias in the absence of magnetic field, has to our knowledge not been observed experimentally.
One possible reason for this is that in SMMs typically $E/D<1$ and the \QK
temperature $T_K$ is suppressed too much, reducing both the height and width of the peak (without splitting it).
\par
{\em Dynamic anisotropy.}
The  anisotropic couplings of the large spin to the vibrational mode $D'$ and $E'$
are however also of importance~\cite{Kortus02},
especially if the vibrational mode frequency energy $\Omega$ is low.
The simplest effect of the longitudinal vibrational coupling $D'$ is a polaronic shift which is different for each magnetic level.
For $E=E'=0$ one can shift the vibrational coordinate $Q$ (or the operator $b$) in $\ham{SMM}$ by an $S_z^2$-dependent amount, resulting in an effective Hamiltonian with eigenvalues
$
  E_{M,n} = -D M^2 -(D'^2/\Omega) M^4  +\Omega n
$, where $M=-S,..,S$ and $n=0,1,..$ are the quantum numbers of the spin and the polaron, respectively.
Each SMM eigenstate is thus dressed by vibrational excitations as sketched in \Fig{fig:1}.
The effective static energy barrier opposing spin inversion is changed in shape and increased in height.
This always results in an effectively \emph{enhanced} ZFS $\delta \rightarrow\Delta$,
\begin{equation}
  \Delta=\delta + (4S(S^2-\frac{3}{2}S+1)-1)\frac{D'^2}{\Omega}
  \label{eq:deltaeff}
  .
\end{equation}
Based on this simple picture one may expect that the \QK effect is suppressed by coupling to vibrations.
\begin{figure}[t!]
   \includegraphics[height=0.5\linewidth]{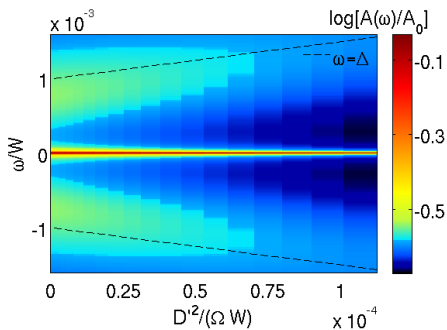}
  \caption{
    (Color online).
    Logarithmic color plot of the SMM spectral function $A(\omega)$
    normalized to the value $A_0=\pi^2/4$ achieved for a symmetrically coupled quantum channel.
    This corresponds to the linear conductance at $\omega=eV$ and $T=0$.
    Parameters $S=3/2$, $D=5\cdot10^{-4}W$, $E=0.1D$, $E'=0$, $\Omega=0.5D$, $J=0.2W$
    and $D'$ is varied.
    The dashed line represents the renormalized ZFS from~\Eq{eq:deltaeff}.
      }
  \label{fig:2}
\end{figure}
For fixed $E'=0$ this is indeed the case as can be seen in \Fig{fig:2}:
Due to the moderate but non-vanishing static $E=0.1D$ a zero-bias \QK peak (red) occurs
which is increasingly suppressed with the coupling $D'$ due to an increased barrier.
In addition two side-peaks are found at the effective ZFS $\omega \approx \pm \Delta$ given by 
\Eq{eq:deltaeff} as indicated by the dashed lines (and renormalized to slightly smaller value due to strong $J$).
These peaks are the remnants of the $S=3/2$ \HK effect.

\begin{figure}[t!]
   \includegraphics[height=0.5\linewidth]{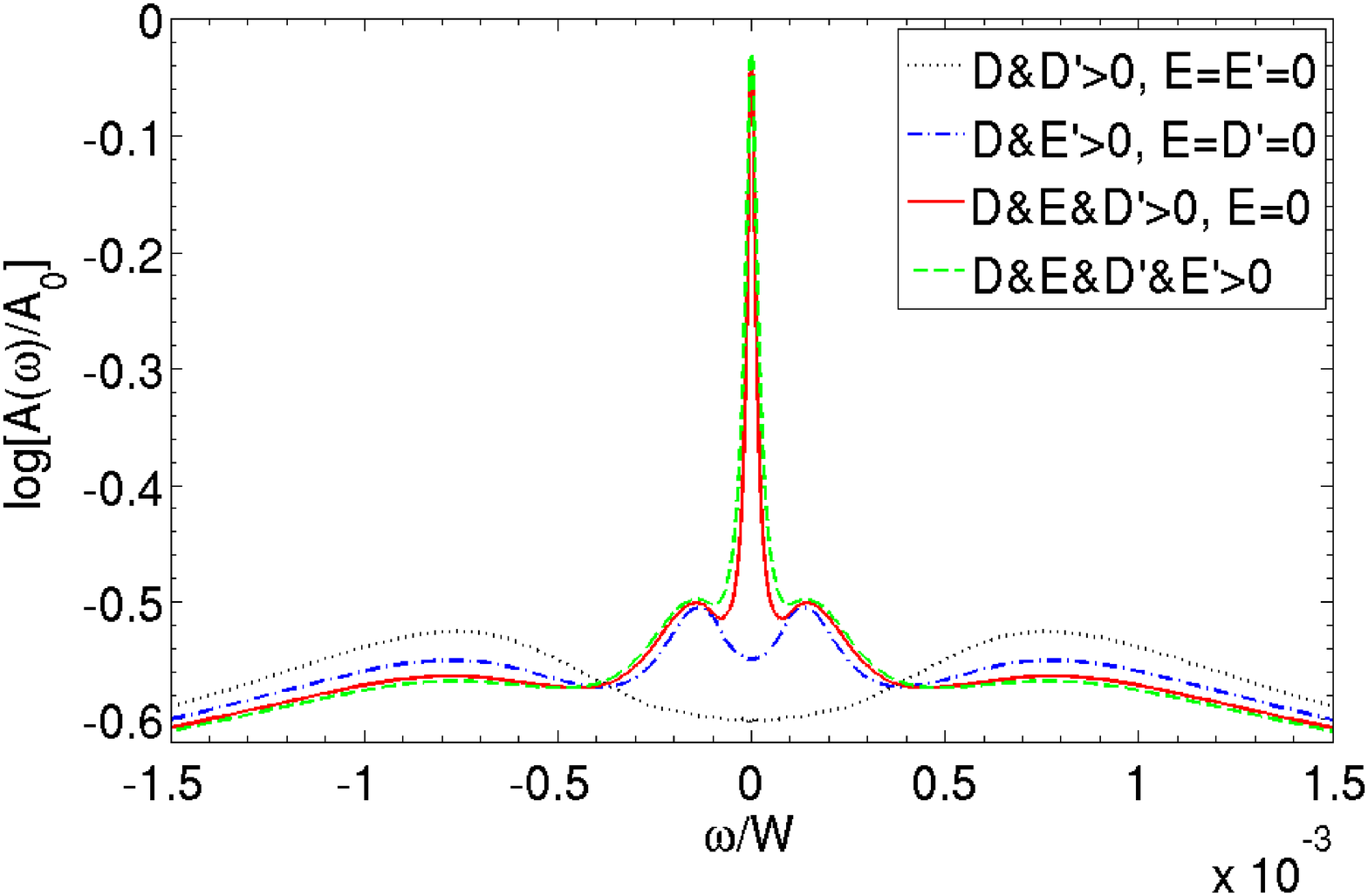}
  \caption{
    (Color online).
    Effect of the spin-vibration coupling on the QST-Kondo peak:
    SMM spectral functions shown for various combinations of zero and non-zero values of the parameters.
    Non-zero values used: $D'=0.04D$, $E'=0.16D$ and $E=0.02D$. Remaining parameters as in \Fig{fig:2}.
  }
  \label{fig:3}
\end{figure}

In \Fig{fig:3} we now focus on the dynamical effect of the spin-vibration coupling
by first considering zero transverse anisotropy ($E=0$).
Without coupling to the vibrations there is no \QK peak.
Switching on spin-vibration coupling of only one type, either $D' \neq 0$ and $E'=0$ (black), or, $E' \neq 0$ and $D'=0$ (blue),
does not change this result.
Only in the latter case, vibrational side-peaks appear in the spectrum at $\omega=\pm \Omega$ (renormalized to smaller value due to strong $J$).
Strikingly, when both types of couplings are non-zero, a pronounced \QK peak appears, even though there is no transverse magnetic anisotropy $E=0$ (red).
This \emph{vibration-induced} \QK effect is the central result of this work.
We now first explain why it requires the presence of both longitudinal and transverse couplings referring to processes sketched in \Fig{fig:1}.
The Kondo effect is related to fluctuations between degenerate states of the SMM which are in opposite Kramer's subspaces~\cite{Romeike06a}.
Since $E=0$, to reach states on the other side of the anisotropy barrier,
 a vibration-induced spin-tunneling of type $E'$ is required, which however,
 involves a virtual vibrational excitation.
An exchange scattering process ($J$), which changes the Kramer's subspace, cannot change the vibrational number at low energy, c.f.~\cite{Paaske05}.
This is why the two processes $E'$ and $J$  result only in a \QK side-peak in \Fig{fig:3} which is split at $\Omega$ and suppressed due to the inability to reach the vibrational ground-state.
Only when a longitudinal spin-coupling $D'$ is present as well, the virtual vibrational excitation can \emph{coherently} reach the ground state and a full \emph{zero-bias} \QK anomaly can develop,
as the red curve in \Fig{fig:3} shows.
One may say that due to the quantum-fluctuations of the vibrational mode of the SMM, the magnetic symmetry is broken in virtual intermediate states, allowing for Kondo exchange scattering, c.f.~\cite{Kikoin06}.
Therefore even for SMMs which have vanishing static $E$ due to symmetry, spin-fluctuations may result in pronounced transport features due to the interplay of exchange scattering and intra-molecular spin-vibration coupling.
Clearly, the vibrational fluctuations can further assist the \QK effect when it already is present due to static $E$:
this results in a higher Kondo temperature as shown in the green curve of \Fig{fig:3}.

\begin{figure}[t!]
   \includegraphics[height=0.4\linewidth]{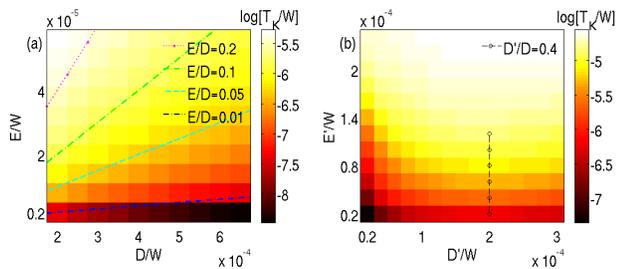}
  \caption{
    (Color online).
    QST-Kondo temperature $T_K$ in units of bandwidth $2W$ (log color scale) determined from the NRG level flow
    for $S=3/2$ and $J=0.2W$.
    (a) Static \QK effect: $T_K$  as function of the static anisotropy parameters $D$, $E$ without spin-vibration coupling.
    Dashed contour lines indicate that  $T_K$ increases with the ratio $E/D$~\cite{Romeike06a}.
    (b) \VQK effect: $T_K$ as function of spin-vibration couplings $D'$ and $E'$ with $\Omega=0.5D$ for 
    static anisotropy, $D=5\cdot10^{-4}W$ and $E=0$.
    The dashed line marks the regime $D'>0.4D$  where $D'$ starts to suppress $T_K$, note the constant $D$ here and also
    the offset: $T_K=0$ for either $E'=0$ or $D'=0$.
    Results for small finite $E$ are similar.
     }
  \label{fig:4}
\end{figure}

In \Fig{fig:4} we show the dependence of the \QK temperature $T_K$ on the longitudinal and transverse static anisotropies and their respective vibration-couplings $D'$ and $E'$.
$T_K$ grows as both the transverse anisotropies are increased since they both enhance QST.
Whereas $D$ always suppresses the \QK effect, \Fig{fig:4}(a), its fluctuations first enhance $T_K$ by allowing QST between the Kramers degenerate ground-states.
Eventually they will suppress $T_K$ if the vibrational contribution to the anisotropy barrier $S^4D'^2/\Omega$, protecting the SMM against these processes, increases too much.

\begin{figure}[t!]
  \includegraphics[height=0.5\linewidth]{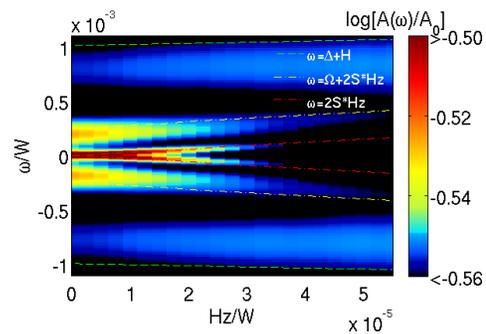}
  \caption{
    (Color online).
    Magnetic field evolution of the spectral function (log color scale) for
    the parameters of the red curve of~\Fig{fig:3}:
    $S=3/2$, $E=0.02D$, $D'=0.04D$ and $E'=0.16D$.
    Dashed lines mark the QST-Kondo (red) and vibrational side-peak (yellow) both evolving with an anomalous g-factor
    and zero-field split high-spin Kondo peak (blue).
  }
  \label{fig:5}
\end{figure}

Finally we show in Fig.~\ref{fig:5} how the experimentally accessible evolution of the conductance with a  magnetic field reveals the different origin of the various peaks.
For simplicity, we consider the field to be along the easy axis as in some experiments~\cite{Zyazin10,Parks10}
and added the Zeeman term $-S_zH_z$ to $\ham{SMM}$ in \Eq{eq1} absorbing the g-factor into the magnetic field.
As the magnetic field is increased,
the \QK peak is weakened and splits with the anomalous $g$-factor, $|\omega_{\text{Kondo}}| = 2 S H_z$~\cite{Romeike11a}.
This clearly indicates the origin of the \QK effect, since the ground-state Kramer's doublet $M=\pm S$ is split by $\Delta M=\pm 2S$.
Strikingly, the vibrational side-peaks have the same, strong field dependence,
 as they correspond to a similar transition offset in energy by $\Omega$:
$|\omega_{\text{vib}}| = \Omega + 2 S H_z$.
In contrast to this, the \HK peak (ZFS) evolves much slower in the magnetic field, independent of the spin magnitude $S$:
$|\omega_{\text{ZFS}}| = \Delta +  H_z$,
signaling that it corresponds only to a transition with $\Delta M=\pm 1$.
Comparing with the above formulas for the peak evolution, indicated by dashed lines in Fig.~\ref{fig:5},
we conclude that the \QK and \HK effects are distinguishable, especially for SMM with large spin.
Finally, we remark that for an SMM with integer spin $S$ there is no \QK effect at zero field, but instead 
a zero-bias conductance dip (for $0<E<D$). As pointed out in~\cite{Leuenberger06} a transverse magnetic field results in a \QK effect,  where similar spin-vibration effects as studied here could occur.

{\em Conclusion.}
We have studied the interplay of spin and vibration on the conductance through a single-molecule magnet.
Whereas longitudinal coupling to the vibration increases the zero-field splitting, suppressing the quantum spin-tunneling Kondo peak,  a vibrationally induced quantum spin-tunneling Kondo effect can occur at zero bias if transverse coupling is present as well.
The transition to virtual vibrational excited states and the transverse spin-mixing in these virtual states
results in a Kondo effect, even in the absence of static transverse anisotropy.
The interplay with vibrations thus can increase the quantum spin-tunneling Kondo temperature for a given static anisotropy, which may motivate further experimental investigation of low temperature transport though single-molecule magnets.
The measurable magnetic field evolution of the conductance
 reveals that vibrational side-bands acquire an anomalous $g$-factor.
We acknowledge A. Cornia, J. Kortus and J. Paaske for stimulating discussions
and support from NanoSci-ERA.
\bibliographystyle{apsrev}

\end{document}